# Multimessenger Astronomy and Astrophysics Synergies

Maurice H.P.M. van Putten[1] and Graziano Rossi

*School of Physics, Korea Institute for Advanced Study, Seoul 130-722, South Korea*

**Executive Summary**

A budget neutral strategy is proposed for NSF to lead the implementation of multimessenger astronomy and astrophysics, as outlined in the Astro2010 Decadal Survey. The emerging capabilities for simultaneous measurements of physical and astronomical data through the different windows of electromagnetic, hadronic and gravitational radiation processes call for a vigorous pursuit of new synergies. The proposed approach is aimed at the formation of new collaborations and multimessenger data-analysis, to transcend the scientific inquiries made within a single window of observations. In view of budgetary constraints, we propose to include the multimessenger dimension in the ranking of proposals submitted under existing NSF programs.

**1. Introduction**

The compelling questions identified in the Astro2010 Decadal Survey[2] concern major thematic areas: the origin of Life and the Universe, cosmic order in the time-dependent Universe, and frontiers of knowledge. The overwhelming message is the emergence of multimessenger astronomy and astrophysics to study the visionary topics mentioned therein, that include but are not limited to extrasolar planets, Dark Matter, structure formation, and the Transient Universe. This development is made possible by advances in technology enabling novel multiwavelength and some radially new multiwindow observations of electromagnetic, hadronic and gravitational radiation processes, as well as advances in resolution and sensitivity in existing windows of observation.

Entirely new methods of discovery are emerging from the prospect of simultaneous measurements of astronomical and physical parameters by which the Universe, in its inception and from the era of reionization to the present, is becoming increasingly accessible to scientific inquiry in the broadest sense of the word. Here, the Transient Universe forms a powerful tracer of key transformative processes, such as supernovae and gamma-ray bursts, the evolution of binaries of neutron stars and black holes, quasars, interacting galaxies and cosmic star formation that, together, are posing some of the most enigmatic questions on the physics of gravity, radiation processes and the evolving Universe.

It is also evident that the sociology of astronomy is changing.[3] Already, the power of discovery by multiwavelength approaches is amply demonstrated, such as by the pioneering Italian-Dutch satellite Beppo-Sax with the first identification of a host galaxy to a cosmological gamma-ray burst.[4] This development gives a continuous push towards building new collaborations across different disciplines and national borders.

Here, we anticipate the need for building new bridges across different windows of observations, to streamline multiwindow data-analysis and to stimulate the development of new strategies for scientific discovery that transcend working in one window alone. We propose an NSF-led initiative on Multimessenger Astronomy and Astrophysics Synergies (MAAS). To vigorously stimulate this development while ensuring a budgetary neutral implementation, we prose that proposals submitted under existing NSF programs are recognized for their MAAS dimension, and receive favorable ranking accordingly.

---

1  Corresponding author, mvputten@kias.re.kr
2  Astro2010 Decadal Survey, 2010, National Academy of Sciences
3  Ref. 1, Chapter 3
4  Shaw Prize Award to Enrico Costa and Gerald Fishman, 2011

A cost-effective realization of the anticipated transformative science, e.g., by the gravitational-wave experiments LIGO, Virgo, and the LCGT, and crucial incremental science, e.g., by the LSST and E-ELT, will require the development of best-practices building on existing experiences and new learning curves. More than ever before, inter-operability and integration of a multitude of data from a diversity of instruments is needed, not unlike ``multi-media" developments in commercial information technology, notably from the following four windows:

  (a) ground based particle detectors, e.g., the Pierre Auger Observatory for observing extragalactic Ultra-High Energy Cosmic Rays (UHECRs), and super-Kamiokande and IceCube for observing low- and high-energy neutrinos;
  (b) the gravitational-wave detectors LIGO, Virgo and the LCGT;
  (c) upcoming advanced electromagnetic observatories such as the LSST and ALMA;
  (d) broad band high-energy satellite observatories, e.g., Fermi/GLAST.

The complexity of the task to maximize the combined strength of these instruments makes it pertinent to develop a structured and systematic approach that is reflected in all echelons of science. As a major funding agency, the NSF is ideally suited to take a leading role in this effort, some of which may be calling for joint operation with NASA, DOE, and eventually ESO and other national and international agencies.

## 2. Specific proposal for building multimessenger synergies

The new instrumentation recommended in the Astro2010 Decadal survey will only be as good as the imagination, skill and organizational talent of their future users. For a budgetary neutral implementation, we propose recognizing proposals submitted to existing NSF programs according to their MAAS dimension with to the following sample criteria in mind, that in any case is expected involve a *combination of at least two distinct windows of observations*:

- Collaborative policies on multimessenger data-analysis. For instance, this may concern MOU's on sharing data, joint publication policies, for the duration of the proposed project or longer;
- Synergies that transcend the power of discovery within a single window of observations. For instance, the development of electromagnetic priors may be invaluable in searches for gravitational-waves from core-collapse supernovae and gamma-ray bursts using novel optical-radio surveys of the local Universe.[5]
- Data-standards to facilitate inter-operability and integration of multimessenger data, with a further objective to reduce time-of-development and to facilitate cost-sharing in data archiving and online data dissemination. These efforts are potentially invaluable in, e.g., creating low-latency pipelines between two different windows, similar to rapid follow-up in the optical of the combined X- and gamm-ray observations pioneered by Beppo-SAX;
- Multimessenger surveys for precision cosmology and probes of the early Universe. For instance, what is an optimal multimessenger approach to probe the nature of inflation?
- Education and training of a new generation of students and post-doctoral researchers in multiwindow astronomy and astrophysics.

## 3. Evaluating MAAS for its impact on science, training and leadership

Possible criteria to evaluate MAAS for its tangible and intangible impact are:

- Creating the kind of "inspiring visibility"[6] that attracts the best and brightest of students and postdoctoral researchers;
- Major discoveries that could not have been realized using one window alone;
- Preserving international leadership n the 21st century;

---

5  van Putten, M.H.P.M., Della Valle, M., & Levinson, A., 2011, A&A Lett., to appear
6  Gerard 't Hooft, 2009, priv. commun.

- Novel best-practices that effectively exploit our multiwindow infrastructure.

## 4. Conclusions

Driven by technological advances, we are on an inevitable course to multimessenger astronomy and astrophysics. How we take advantage of this development, in scientific yield, cost-effectiveness and leadership, depends crucially on vision, imagination and management of new collaborations that transcend our current practices. As a major funding agency, NSF can take a leading role in this exciting development.

We propose a budget neutral strategy for the realization of multimessenger astronomy and astrophysics with no change in the priorities set forth in the Astro2010 Decadal Survey, by recognizing the multimessenger dimension in the ranking of proposals submitted to existing NSF programs. Some sample criteria for MAAS and its intended impact on research, education and international leadership are mentioned for future consideration.